\documentclass[journal]{IEEEtran}

\usepackage[noadjust]{cite}

\usepackage[cmex10]{amsmath}
\usepackage{array}
\usepackage{amssymb}
\usepackage{multirow}

\newcommand{\eqdef}{\overset{\mathrm{def}}{=}}

\newskip\smallskipamount % the amount of a \smallskip
\def\smallskip{\vskip\smallskipamount}

\usepackage[usenames,dvipsnames]{pstricks}
\usepackage{epsfig}
\usepackage{threeparttable}

\usepackage[ruled,norelsize]{algorithm2e}

\makeatletter
\newcommand{\removelatexerror}{\let\@latex@error\@gobble}
\makeatother

\begin{document}
\title{A High-Throughput Energy-Efficient Implementation of Successive Cancellation Decoder for Polar Codes Using Combinational Logic}
\author{Onur~Dizdar,~\IEEEmembership{Student Member,~IEEE,}
        and~Erdal~Ar\i kan,~\IEEEmembership{Fellow,~IEEE}%
\thanks{Copyright (c) 2015 IEEE. Personal use of this material is permitted. However, permission to use this material for any other purposes must be obtained from the IEEE by
sending an email to pubs-permissions@ieee.org.}%
\thanks{The authors are with the Department of Electrical-Electronics Engineering, Bilkent University, Ankara, TR-06800, Turkey 
(e-mail: \mbox{odizdar@ee.bilkent.edu.tr}, arikan@ee.bilkent.edu.tr.)}}%
\maketitle

\begin{abstract}
This paper proposes a high-throughput energy-efficient Successive Cancellation (SC) decoder architecture for polar codes based on combinational logic. 
The proposed combinational architecture operates at relatively low clock frequencies compared to sequential circuits, 
but takes advantage of the high degree of parallelism inherent in such architectures to provide
a favorable tradeoff between throughput and energy efficiency at short to medium block lengths. At longer block lengths, the paper proposes a hybrid-logic SC decoder that combines the advantageous aspects of the combinational decoder with the low-complexity nature of sequential-logic decoders. Performance characteristics on ASIC and FPGA are presented with a detailed power consumption analysis for combinational decoders. Finally, the paper presents an analysis of the complexity and delay of combinational decoders, and of the throughput gains obtained by hybrid-logic decoders with respect to purely synchronous architectures.
\end{abstract}


\IEEEpeerreviewmaketitle

\section{Introduction}
\IEEEPARstart{P}{olar} codes were proposed in \cite{arikan} as a low-complexity channel coding method that can provably achieve Shannon's channel capacity for any binary-input symmetric discrete memoryless channel. Apart from the intense theoretical interest in the subject, polar codes have attracted attention for their potential applications. There have been several proposals on hardware implementations of polar codes, which mainly focus on maximizing throughput or minimizing hardware complexity. In this work, we propose an architecture for SC decoding using combinational logic in an effort to obtain a high throughput decoder with low power consumption. We begin with a survey of the relevant literature.

The basic decoding algorithm for polar codes is the SC decoding algorithm, which is a non-iterative sequential algorithm with complexity $O(N\log N)$ for a code of length $N$. 
Many of the SC decoding steps can be carried out in parallel and the latency of the SC decoder can be reduced to roughly $2N$ in a fully-parallel implementation, as pointed out in \cite{arikan}
and \cite{Arikan2010}.
This means that the throughput of any synchronous SC decoder is limited to $\frac{f_{c}}{2}$ in terms of the clock frequency $f_{c}$, as pointed out in \cite{hardwarearchitectures}. 
The throughput is reduced further in semi-parallel architectures, such as \cite{scasic} and \cite{anefficientpartsumnet}, which increase the decoding latency further in exchange for reduced hardware complexity. This throughput bottleneck in SC decoding is inherent in the logic of SC decoding and stems from the fact that the decoder makes its final decisions one at a time in a sequential manner.

Some algorithmic and hardware implementation methods have been proposed to overcome the throughput bottleneck problem in polar decoding. One method that has been tried is Belief Propagation (BP) decoding, starting with \cite{arikanbp}. 
In BP decoding, the decoder has the capability of making multiple bit decisions in parallel. 
Indeed, BP throughputs of $2$~Gb/s (with clock frequency $500$~MHz) and $4.6$~Gb/s (with clock frequency $300$~MHz) are reported in \cite{architecturesforpolarbp} and \cite{bpasicthesis}, respectively. 
Generally speaking, the throughput advantage of BP decoding is observed at high SNR values, where correct decoding can be achieved after a small number of iterations; this advantage of BP decoders over SC decoders diminishes as the SNR decreases.

A second algorithmic approach to break the throughput bottleneck is to exploit the fact that polar codes are a class of generalized concatenated codes (GCC). More precisely, a polar code $\mathcal{C}$ of length-$N$ is constructed from two length-$N/2$ codes $\mathcal{C}_{1}$ and $\mathcal{C}_{2}$, using the well-known Plotkin $|\mathbf{u}|\mathbf{u}+\mathbf{v}|$ code combining technique \cite{Plotkin}. 
The recursive nature of the polar code construction ensures that the constituent codes $\mathcal{C}_1$ and $\mathcal{C}_2$ are polar codes in their own right and each can be further decomposed into two polar codes of length $N/4$, and so on, until the block-length is reduced to one. In order to improve the throughput of a polar code, one may introduce specific measures to speed up the decoding of the constituent polar codes encountered in the course of such recursive decomposition. For example, when a constituent code $\mathcal{C}_{i}$ of rate $0$ or $1$ is encountered, the decoding becomes a trivial operation and 
can be completed in one clock cycle. Similarly, decoding is trivial when the constituent code is a repetition code or a single parity-check code.
Such techniques have been applied earlier in the context of Reed-Muller codes by \cite{schnabl_bossert} and \cite{dumer_shabunov}. They have been also used in speeding up SC decoders for polar codes by \cite{kschis}. Results reported by such techniques show a throughput of $1$ Gb/s by using designs tailored for specific codes \cite{fastpolardecoders}. On the other hand, decoders utilizing such shortcuts require reconfiguration when the code is changed, which makes their use difficult in systems using adaptive coding methods.

Implementation methods such as precomputations, pipelined, and unrolled designs, have also been proposed to improve the throughput of SC decoders. These methods trade hardware complexity for gains in throughput. For example, it has been shown that the decoding latency may be reduced to $N$ by doubling the number of adders in a SC decoder circuit \cite{lowlatencysequential}. 
A similar approach has been used in a first ASIC implementation of a SC decoder to reduce the latency at the decision-level LLR calculations by $N/2$ clock cycles and provide a throughput of $49$~Mb/s with $150$~MHz clock frequency for a rate-$1/2$ code \cite{scasic}. In contrast, pipelined and unrolled designs do not affect the latency of the decoder; the increase in throughput is obtained by decoding multiple codewords simultaneously without resource sharing. A recent study \cite{unrolled_final} exhibits a SC decoder achieving $254$ Gb/s throughput with a fully-unrolled and deeply-pipelined architecture using component code properties for a rate-$1/2$ code. Pipeling in the context of polar decoders was used earlier in various forms and in a more limited manner in \cite{Arikan2010}, \cite{hardwarearchitectures}, \cite{Pamuk2011}, \cite{lowlatencysequential}, and \cite{interleavedsc}.
 
SC decoders, while being simple, are suboptimal. In \cite{tal_list}, SC {\sl list-of-$L$} decoding was proposed for decoding polar codes, following similar ideas 
developed earlier by \cite{DumerList} for Reed-Muller codes.
Ordinary SC decoding is a special case of SC list decoding with list size $L=1$.
SC list decoders show markedly better performance compared to SC decoders at the expense of complexity, and are subject to the same throughput bottleneck problems as ordinary SC decoding.
Parallel decision-making techniques, as discussed above, can be applied to improve the throughput of SC list decoding.
For instance, it was shown in \cite{parhi_list} that by using $4$-bit parallel decisions, a list-of-2 SC decoder can achieve a throughput of around $500$~Mb/s with a clock frequency of $500$~MHz.

The present work is motivated by the desire to obtain high-throughput SC decoders with low power consumption, which has not been a main concern in literature so far. These desired properties are attained by designing completely combinational decoder architectures, which is possible thanks to the recursive and feed-forward (non-iterative) structure of the SC algorithm. Combinational decoders operate at lower clock frequencies compared to ordinary synchronous (sequential logic) decoders. However, in a combinational decoder an entire codeword is decoded in one clock cycle. This allows combinational decoders to operate with less power while maintaining a high throughput, as we demonstrate in the remaining sections of this work. 

Pipelining can be applied to combinational decoders at any depth to adjust their throughput, hardware usage, and power consumption characteristics. Therefore, we also investigate the performance of pipelined combinational decoders. We do not use any of the multi-bit decision shortcuts in the architectures we propose. 
Thus, for a given block length, the combinational decoders that we propose retain the inherent flexibility of polar coding to operate at any desired code rate between zero and one. 
Retaining such flexibility is important since one of the main motivations behind the combinational decoder is to use it as an ``accelerator'' module as part of a hybrid decoder that combines a synchronous SC decoder with a combinational decoder to take advantage of the best characteristics of the two types of decoders. 
We give an analytical discussion of the throughput of hybrid-logic decoders to quantify the advantages of the hybrid decoder.

The rest of this paper is organized as follows.
Section~\ref{sec:background} give a brief discussion of polar coding to define the SC decoding algorithm. 
Section~\ref{sec:combinational} introduces the main decoder architectures considered in this paper, namely, combinational decoders, pipelined combinational decoders, and hybrid-logic decoders.
Also included in that section is an analysis of the hardware complexity and latency of the proposed decoders. 
Implementation results of combinational decoders and pipelined combinational decoders are presented in Section~\ref{sec:implementation}, with a detailed power consumption analysis for combinational decoders. Also presented in the same section is an analysis of the throughput improvement obtained by hybrid-logic decoders relative to synchronous decoders. 
Section~\ref{sec:conclusion} concludes the paper.

Throughout the paper, vectors are denoted by boldface lowercase letters. All matrix and vector operations are over vector spaces over the binary field $\mathbb{F}_{2}$. Addition over $\mathbb{F}_{2}$ is represented by the $\oplus$ operator. For any set $\mathcal{S} \subseteq \left\{0,1,\ldots, N-1\right\}$, $\mathcal{S}^{\mathrm{c}}$ denotes its complement. For any vector $\mathbf{u}=\left(u_{0}, u_{1},\ldots , u_{N-1}\right)$ of length $N$ and set $\mathcal{S} \subseteq \left\{0,1,\ldots, N-1\right\}$,  \mbox{$\mathbf{u}_{\mathcal{S}}\eqdef\left[u_{i} \ : i \in \mathcal{S} \right]$}. We define a binary sign function $\mathrm{s}(\ell)$ as
\begin{align}
\mathrm{s}(\ell)=
\left\{
	\begin{array}{ll}
		0,  & \mbox{if} \ \ell \geq 0 \\
		1, & \mbox{otherwise}.
	\end{array}
\right.
\label{eq:sign}
\end{align} 

\section{Background on Polar Coding}
\label{sec:background}
\begin{figure}[h!]
   \centering
\vspace{-0.4cm}
\includegraphics[width=6.3in,height=6.3in,keepaspectratio]{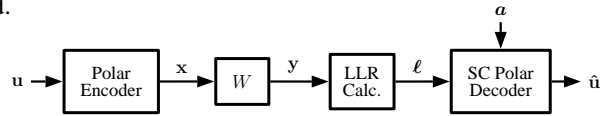}
\vspace{-1.3cm}
\caption{Communication scheme with polar coding} 
\label{fig:commn}
\end{figure}

We briefly describe the basics of polar coding in this section, including the SC decoding algorithm. Consider the system given in Fig.~\ref{fig:commn}, in which a polar code is used for channel coding. 
All input/output signals in the system are vectors of length $N$, where $N$ is the length of the polar code that is being used.

The encoder input vector \mbox{$\mathbf{u} \in \mathbb{F}_{2}^{N}$} consists of a {\sl data} part $\mathbf{u}_{\mathcal{A}}$ and a {\sl frozen} part $\mathbf{u}_{\mathcal{A}^{\mathrm{c}}}$, where $\mathcal{A}$ is chosen in accordance with polar code design rules as explained in \cite{arikan}. 
We fix the frozen part $\mathbf{u}_{\mathcal{A}^{\mathrm{c}}}$ to zero in this study. 
We define a {\sl frozen-bit indicator vector} $\boldsymbol{a}$ so that $\boldsymbol{a}$ is a 0-1 vector of length $N$ with
\begin{align}
a_{i}=
\left\{
	\begin{array}{ll}
		0,  & \mbox{if} \ i \in \mathcal{A}^{\mathrm{c}} \\  \nonumber
		1,  & \mbox{if} \ i \in \mathcal{A} .  \nonumber
	\end{array}
\right.  \nonumber
\end{align}
The frozen-bit indicator vector is made available to the decoder in the system.

The channel $W$ in the system is an arbitrary discrete memoryless channel with input alphabet ${\cal X}=\{0,1\}$, output alphabet ${\cal Y}$ and transition probabilities $\{W(y|x):x\in {\cal X},y\in {\cal Y}\}$.
In each use of the system, a codeword $\mathbf{x} \in \mathbb{F}_{2}^{N}$ is transmitted, and a channel output vector \mbox{$\mathbf{y} \in \mathcal{Y}^{N}$} is received. 
The receiver calculates a log-likelihood ratio (LLR) vector ${\mathbf \ell}=(\ell_1,\ldots,\ell_N)$ with
$$
\ell_i = \ln \left(\frac{P\left(y_{i} | x_{i}=0\right)}{P\left(y_{i} | x_{i}=1\right)}\right),
$$
and feeds it into the SC decoder.

\begin{figure}[h!]
 \removelatexerror
  \begin{algorithm}[H]
	 \caption{$\mathbf{\hat{u}}=\textsc{Decode}(\boldsymbol{\ell}, \boldsymbol{a})$}
		\label{alg:2NbyN_dec}
			$N=$\textit{length}$(\boldsymbol{\ell})$ \\
			\eIf{$N==2$}{
			$\hat{u}_{0} \gets \mathrm{s}\left(f(\ell_{0},\ell_{1})\right)\cdot a_{0}$\\
			$\hat{u}_{1} \gets \mathrm{s}\left(g(\ell_{0},\ell_{1},\hat{u}_{0})\right)\cdot a_{1}$\\
			\Return $\mathbf{\hat{u}} \gets (\hat{u}_{0}, \hat{u}_{1})$
			}{
			$\boldsymbol{\ell}^{\prime} \gets f_{N/2}(\boldsymbol{\ell})$																				 \\
			$\boldsymbol{a}^{\prime} \gets (a_{0}, \ldots, a_{N/2-1})$																						   \\
			$\mathbf{\hat{u}}^{\prime} \gets \textsc{Decode}(\boldsymbol{\ell}^{\prime}, \boldsymbol{a}^{\prime})$  \\
			$\mathbf{v} \gets \textsc{Encode}(\mathbf{\hat{u}}^{\prime})$    																	\\
			$\boldsymbol{\ell}^{\prime \prime}\gets g_{N/2}(\boldsymbol{\ell}, \mathbf{v})$												\\
			$\boldsymbol{a}^{\prime \prime} \gets (a_{N/2}, \ldots, a_{N-1})$																			\\
			$\mathbf{\hat{u}}^{\prime \prime}\gets \textsc{Decode}(\boldsymbol{\ell}^{\prime \prime}, \boldsymbol{a}^{\prime \prime})$  \\
			\Return $\mathbf{\hat{u}} \gets (\mathbf{\hat{u}}^{\prime}, \mathbf{\hat{u}}^{\prime \prime})$         
		}
  \end{algorithm}
\end{figure}

The decoder in the system is an SC decoder as described in \cite{arikan}, which takes as input the channel LLRs and the frozen-bit indicator vector and calculates an estimate \mbox{$\mathbf{\hat{u}}  \in \mathbb{F}_{2}^{N}$} of the data vector $\mathbf{u}$. The SC algorithm outputs bit decisions sequentially, one at a time in natural index order, with each bit decision depending on prior bit decisions. 
A precise statement of the SC algorithm is given in Algorithm~\ref{alg:2NbyN_dec},
where the functions $f_{N/2}$ and $g_{N/2}$ are defined as
\begin{gather*}
 f_{N/2}(\boldsymbol{\ell})=\left(f(\ell_{0},\ell_{1}), \ldots, f(\ell_{N-2},\ell_{N-1})\right)\\ 
 g_{N/2}(\boldsymbol{\ell}, \mathbf{v})=\left(g(\ell_{0},\ell_{1},v_{0}), \ldots, g(\ell_{N-2},\ell_{N-1},v_{N/2-1})\right)
\end{gather*} 
with
\begin{gather*}
 f(\ell_{1},\ell_{2})=2\tanh^{-1}\left(\tanh\left(\ell_1/2\right) \,\tanh\left(\ell_2/2\right) \right) \\
 g(\ell_{1},\ell_{2},v)=\ell_{1}(-1)^{v} + \ell_{2}.
\end{gather*}  
In actual implementations discussed in this paper, the function $f$ is approximated using the {\sl min-sum} formula
\begin{equation}
f(\ell_{1},\ell_{2})\approx(1-2\mathrm{s}(\ell_{1}))\cdot (1-2\mathrm{s}(\ell_{2}))\cdot \min\left\{\left|\ell_{1}\right|,\left|\ell_{2}\right|\right\},
\label{eq:f_minsum}
\end{equation}
and $g$ is realized in the alternative (exact) form
\begin{equation}
 g(\ell_{1},\ell_{2},v)=\ell_{2}+(1-2v)\cdot \ell_{1}.
\label{eq:g_minsum}
\end{equation}
   
A key property of the SC decoding algorithm that makes low-complexity implementations possible is its recursive nature, where a decoding instance of block length $N$ is broken in the decoder into two decoding instances of lengths $N/2$ each. 

\section{SC Decoder Using Combinational Logic}
\label{sec:combinational}
The pseudocode in Algorithm~\ref{alg:2NbyN_dec} shows that the logic of SC decoding contains no loops, hence it can be implementated using only combinational logic. 
The potential benefits of a combinational implementation are high throughput and low power consumption, which we show are feasible goals. 
In this section, we first describe a combinational SC decoder for length $N=4$ to explain the basic idea.
Then, we describe the three architectures that we propose. Finally, we give an analysis of complexity and latency 
characteristics of the proposed architectures.
\subsection{Combinational Logic for SC Decoding}
\begin{figure}[h!]
%\centering
\hspace{0.8cm}\includegraphics[width=2.5in,height=2.5in,keepaspectratio]{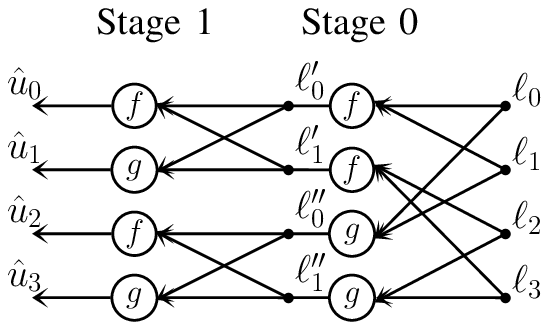}
\caption{SC decoding trellis for $N=4$}
\label{fig:dec_graph4}
\end{figure}

In a combinational SC decoder the decoder outputs are expressed directly in terms of decoder inputs, without any registers or memory elements in between the input and output stages.
Below we give the combinational logic expressions for a decoder of size $N=4$, for which the signal flow graph (trellis) is depicted in Fig.~\ref{fig:dec_graph4}.

At Stage 0 we have the LLR relations
\begin{gather*}
\ell_{0}^{\prime}=f(\ell_{0},\ell_{1}), \quad \ell_{1}^{\prime}=f(\ell_{2},\ell_{3}),\\
\ell_{0}^{\prime\prime}=g(\ell_{0},\ell_{1},\hat{u}_{0} \oplus \hat{u}_{1}),\quad 
\ell_{1}^{\prime\prime}=g(\ell_{2},\ell_{3},\hat{u}_{1}).
\end{gather*} 
At Stage 1, the decisions are extracted as follows.
\begin{gather*}
\hat{u}_{0}=\mathrm{s}\left[f\left(f(\ell_{0},\ell_{1}), f(\ell_{2},\ell_{3})\right)\right]\cdot a_{0}, \\
\hat{u}_{1}=\mathrm{s}\left[g\left(f(\ell_{0},\ell_{1}), f(\ell_{2},\ell_{3}),\hat{u}_{0}\right)\right]\cdot a_{1},\\
\hat{u}_{2}=\mathrm{s}\left[f\left(g(\ell_{0},\ell_{1},\hat{u}_{0} \oplus \hat{u}_{1}),g(\ell_{2},\ell_{3},\hat{u}_{1})\right)\right]\cdot a_{2},\\
\hat{u}_{3}=\mathrm{s}\left[g\left(g(\ell_{0},\ell_{1},\hat{u}_{0} \oplus \hat{u}_{1}),g(\ell_{2},\ell_{3},\hat{u}_{1}),\hat{u}_{2}\right)\right]\cdot a_{3},
\end{gather*} 
where the decisions $\hat{u}_0$ and $\hat{u}_2$ may be simplified as
\begin{gather*}
\hat{u}_{0}=\left[\mathrm{s}(\ell_{0}) \oplus \mathrm{s}(\ell_{1}) \oplus \mathrm{s}(\ell_{2}) \oplus \mathrm{s}(\ell_{3})\right]\cdot a_{0}, \\
\hat{u}_{2}=\left[\mathrm{s}\left(g(\ell_{0},\ell_{1},\hat{u}_{0} \oplus \hat{u}_{1})\right) \oplus \mathrm{s}\left(g(\ell_{2},\ell_{3},\hat{u}_{1})\right)\right]\cdot a_{2}.
\end{gather*} 

\begin{figure}[h!]
   %\centering
	\includegraphics[width=3.5in,height=3.5in,keepaspectratio]{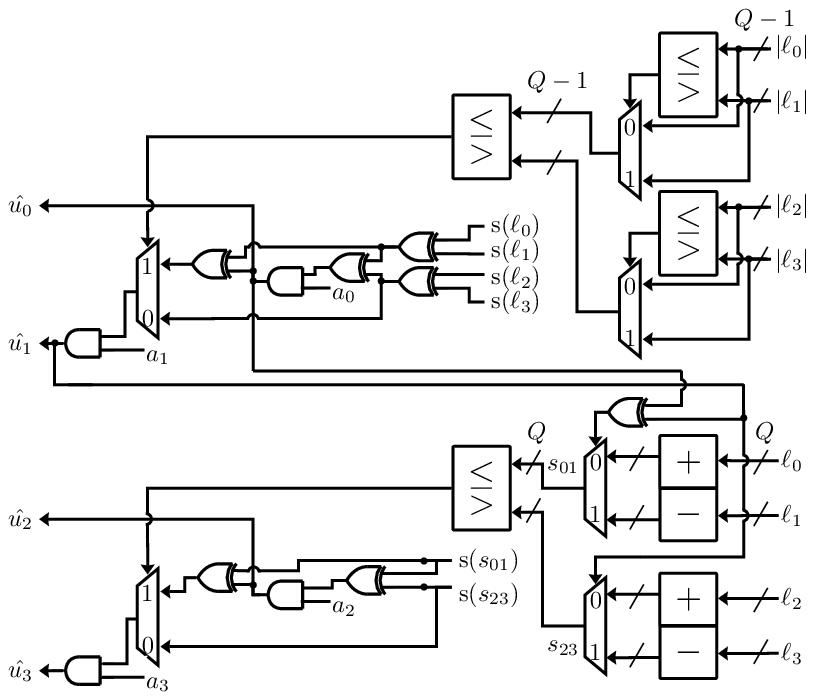}
	\vspace{-0.5cm}
\caption{Combinational decoder for $N=4$} 
\label{fig:N4decoder}
\end{figure} 
\label{sec:combdec}

\begin{figure*}[ht!]
   %\centering
		\hspace{1.0cm}\includegraphics[width=8.0in,height=8.0in,keepaspectratio]{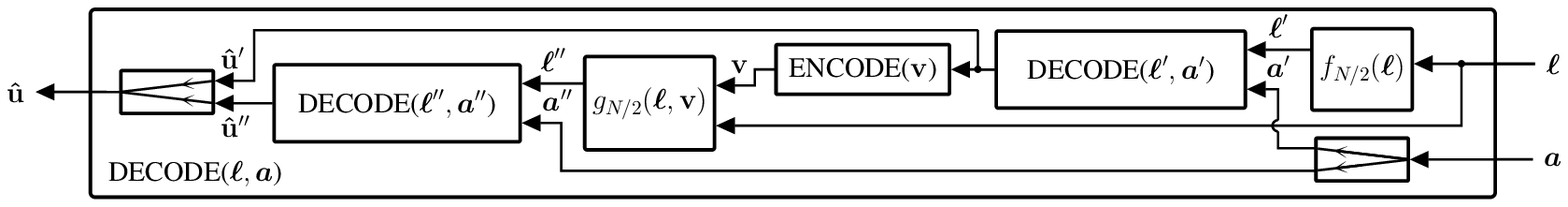}
		\vspace{-3.2cm}
\caption{Recursive architecture of polar decoders for block length $N$} 
\label{fig:2NbyN_fig}
\end{figure*} 

Fig.~\ref{fig:N4decoder} shows a combinational logic implementation of the above decoder using only comparators and adders.
We use sign-magnitude representation, as in \cite{asemiparallel}, to avoid excessive number of conversions between different representations.
Channel observation LLRs and calculations throughout the decoder are represented by $Q$ bits. The function $g$ of \eqref{eq:g_minsum} is implemented using the precomputation method
suggested in \cite{lowlatencysequential} to reduce latency. 
In order to reduce latency and complexity further, we implement the decision logic for odd-indexed bits as
\begin{align}
\hat{u}_{2i+1}=
\left\{ 
	\begin{array}{ll}
		0\hfill,&\ \mbox{if}  \ a_{2i+1} = 0  \\
		\mathrm{s}(\lambda_{2})\hfill,& \ \mbox{if}  \ a_{2i+1} = 1 \ \mbox{and} \ \left|\lambda_{2}\right| \geq \left|\lambda_{1}\right| \\
		\mathrm{s}(\lambda_{1}) \oplus \hat{u}_{2i},& \ \mbox{otherwise}.  
	\end{array}
\right.  
\label{eq:compinsteadofadd}
\end{align} 

\subsection{Architectures}

In this section, we propose three SC decoder architectures for polar codes: combinational, pipelined combinational, and hybrid-logic decoders. 
Thanks to the recursive structure of the SC decoder, the above combinational decoder of size $N=4$ will serve as a basic building block for the larger decoders that we discuss
in the next subsection.

\subsubsection{Combinational Decoder}

A combinational decoder architecture for any block length $N$ using the recursive algorithm in Algorithm~\ref{alg:2NbyN_dec} is shown in Fig.~\ref{fig:2NbyN_fig}.
This architecture uses two combinational decoders of size $N/2$, with glue logic consisting of one $f_{N/2}$ block, one $g_{N/2}$ block, and one size-$N/2$ encoder block.

\begin{figure}[h!]
   %\centering
			\hspace{0.9cm}\includegraphics[width=5.75in,height=5.75in,keepaspectratio]{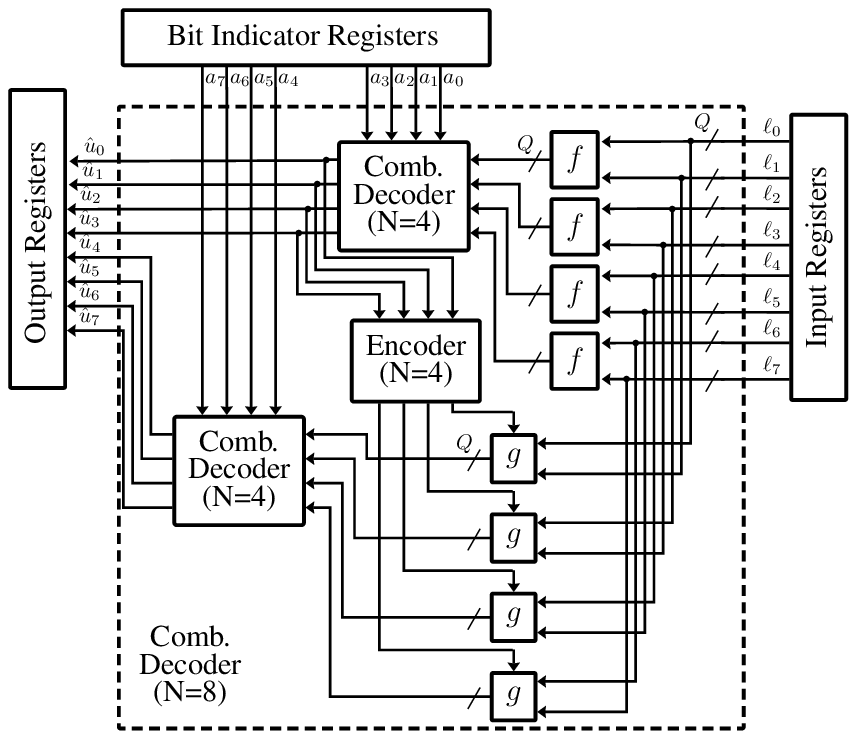}
			\vspace{-2.4cm}
\caption{RTL schematic for combinational decoder ($N=8$)} 
\label{fig:N8decoder}
\end{figure} 

The RTL schematic for a combinational decoder of this type is shown in Fig.~\ref{fig:N8decoder} for $N=8$.   
The decoder submodules of size-4 are the same as in Fig.~\ref{fig:N4decoder}. 
The size-4 encoder is implemented using combinational circuit consisting of XOR gates. 
The logic blocks in a combinational decoder are directly connected without any synchronous logic elements in-between, which
helps the decoder to save time and power by avoiding memory read/write operations. 
Avoiding the use of memory also reduces hardware complexity.
In each clock period, a new channel observation LLR vector is read from the input registers and a decision vector is written to the output registers. 
The clock period is equal to the overall combinational delay of the circuit, which determines the throughput of the decoder. 
The decoder differentiates between frozen bits and data bits by AND gates and the frozen bit indicators $a_{i}$, as shown in Fig.~\ref{fig:N4decoder}. The frozen-bit indicator vector 
can be changed at the start of each decoding operation, making it possible to change the code configuration in real time.
Advantages and disadvantages of combinational decoders will be discussed in more detail in Section~\ref{sec:implementation}.

\subsubsection{Pipelined Combinational Decoder}
\label{sec:pipelined}
Unlike sequential circuits, the combinational architecture explained above has no need for any internal storage elements. 
The longest path delay determines the clock period in such a circuit.
This saves hardware by avoiding usage of memory, but slows down the decoder.
In this subsection, we introduce pipelining in order to increase the throughput at the expense of some extra hardware utilization.

It is seen in Fig.~\ref{fig:2NbyN_fig} that the outputs of the first decoder block (DECODE($\boldsymbol{\ell}^{\prime}, \boldsymbol{a}^{\prime}$)) are used by the encoder to calculate partial-sums. Therefore, this decoder needs to preserve its outputs after they settle to their final values. However, this particular decoder can start the decoding operation for another codeword if these partial-sums are stored with the corresponding channel observation LLRs for the second decoder (DECODE($\boldsymbol{\ell}^{\prime\prime}, \boldsymbol{a}^{\prime\prime}$)). Therefore, adding register blocks to certain locations in the decoder enable a pipelined decoding process. 

Early examples of pipelining in the context of synchronous polar decoders are \cite{Arikan2010}, \cite{hardwarearchitectures}, and \cite{Pamuk2011}.
In synchronous design with pipelining, shared resources at certain stages of decoding have to be duplicated in order to prevent conflicts on calculations when multiple codewords are processed in the decoder. The number of duplications and their stages depend on the number of codewords to be processed in parallel. Since pipelined decoders are derived from combinational decoders, they do not use resource sharing; therefore, resource duplications are not needed. Instead, pipelined combinational decoders aim to reuse the existing resources. This resource reuse is achieved by using storage elements to save the outputs of smaller combinational decoder components and re-employ them in decoding of another codeword. 
\begin{figure*}[hbt]
   %\centering
		\hspace{3.0cm}\includegraphics[width=7.0in,height=7.0in,keepaspectratio]{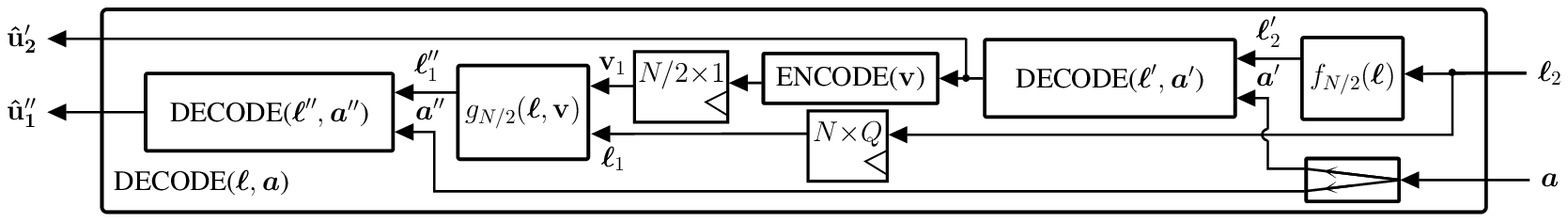}
		\vspace{-3.1cm}
\caption{Recursive architecture for pipelined polar decoders for block length $N$} 
\label{fig:2NbyN_pipe_fig}
\end{figure*} 

A single stage pipelined combinational decoder is shown in Fig.~\ref{fig:2NbyN_pipe_fig}. The channel observation LLR vectors $\boldsymbol{\ell}_{1}$ and $\boldsymbol{\ell}_{2}$ in this architecture correspond to different codewords. The partial-sum vector $\mathbf{v}_{1}$ is calculated from the first half of the decoded vector for $\boldsymbol{\ell}_{1}$. Output vectors $\mathbf{\hat{u}}_{2}^{\prime}$  and $\mathbf{\hat{u}}_{1}^{\prime \prime}$ are the first and second halves of decoded vectors for $\boldsymbol{\ell}_{2}$ and $\boldsymbol{\ell}_{1}$, respectively. The schedule for this pipelined combinational decoder is given in Table~\ref{table:pipelinedschedule}.  

\begin{table}[hbt!]
\caption{Schedule for Single Stage Pipelined Combinational Decoder} % title of Table
\centering % used for centering table
\begin{tabular}{>{\centering\arraybackslash} m{0.8in} | >{\centering\arraybackslash} m{0.1in} | >{\centering\arraybackslash} m{0.1in} | >{\centering\arraybackslash} m{0.1in} | >{\centering\arraybackslash} m{0.1in} | >{\centering\arraybackslash} m{0.1in} | >{\centering\arraybackslash} m{0.1in} | >{\centering\arraybackslash} m{0.1in} | >{\centering\arraybackslash} m{0.1in}} % centered columns (4 columns)
\hline %inserts double horizontal lines
Clock Cycle & 1 & 2 & 3 & 4 & 5 & 6 & 7 & 8  \\  [0.5ex] % inserts table 
%heading
\hline\hline % inserts single horizontal line
Input of DECODE($\boldsymbol{\ell}, \boldsymbol{a}$) & $\boldsymbol{\ell}_{1}$ & $\boldsymbol{\ell}_{2}$ & $\boldsymbol{\ell}_{3}$ & $\boldsymbol{\ell}_{4}$ & $\boldsymbol{\ell}_{5}$ & $\boldsymbol{\ell}_{6}$ &   \\  \hline
Output of DECODE($\boldsymbol{\ell}^{\prime}, \boldsymbol{a}^{\prime}$) &  & $\mathbf{\hat{u}_{1}}^{\prime}$ & $\mathbf{\hat{u}_{2}}^{\prime}$ & $\mathbf{\hat{u}_{3}}^{\prime}$ & $\mathbf{\hat{u}_{4}}^{\prime}$ & $\mathbf{\hat{u}_{5}}^{\prime}$ & $\mathbf{\hat{u}_{6}}^{\prime}$ &    \\  \hline
Output of DECODE($\boldsymbol{\ell}^{\prime \prime}, \boldsymbol{a}^{\prime \prime}$)  &  &  & $\mathbf{\hat{u}_{1}}^{\prime \prime}$ & $\mathbf{\hat{u}_{2}}^{\prime \prime}$ & $\mathbf{\hat{u}_{3}}^{\prime \prime}$ & $\mathbf{\hat{u}_{4}}^{\prime \prime}$ & $\mathbf{\hat{u}_{5}}^{\prime \prime}$ & $\mathbf{\hat{u}_{6}}^{\prime \prime}$  \\   \hline
Output of DECODE($\boldsymbol{\ell}, \boldsymbol{a}$) & & & $\mathbf{\hat{u}_{1}}$ & $\mathbf{\hat{u}_{2}}$ & $\mathbf{\hat{u}_{3}}$ & $\mathbf{\hat{u}_{4}}$ & $\mathbf{\hat{u}_{5}}$ & $\mathbf{\hat{u}_{6}}$     \\   \hline
%\hline %inserts single line
\end{tabular}
\label{table:pipelinedschedule} % is used to refer this table in the text
\end{table} 

As seen from Table~\ref{table:pipelinedschedule}, pipelined combinational decoders, like combinational decoders, decode one codeword per clock cycle. However, the maximum path delay of a pipelined combinational decoder for block length $N$ is approximately equal to the delay of a combinational decoder for block length $N/2$. Therefore, the single stage pipelined combinational decoder in Fig.~\ref{fig:2NbyN_pipe_fig} provides approximately twice the throughput of a combinational decoder for the same block length. On the other hand, power consumption and hardware usage increase due to the added storage elements and increased operating frequency. Pipelining stages can be increased by making the two combinational decoders for block length $N/2$ in Fig.~\ref{fig:2NbyN_pipe_fig} also pipelined in a similar way to increase the throughput further. Comparisons between combinational decoders and pipelined combinational decoders are given in more detail in Section~\ref{sec:implementation}.

\subsubsection{Hybrid-Logic Decoder}
\label{sec:hybrid}
\begin{figure}[h!]
%\centering
		\hspace{1.4cm}\includegraphics[width=2.6in,height=2.6in,keepaspectratio]{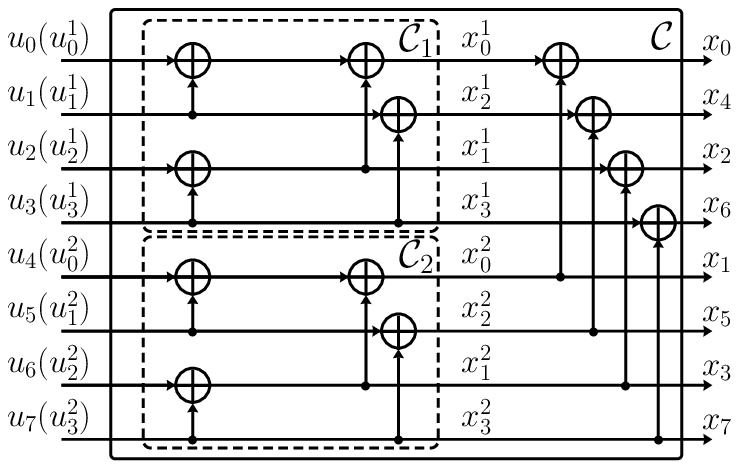}
		\vspace{0.1cm}
\caption{Encoding circuit of $\mathcal{C}$ with component codes $\mathcal{C}_{1}$ and $\mathcal{C}_{2}$ ($N=8$ and $N^{\prime}=4$)}
\label{fig:hybrid_enc}
\end{figure}

In this part, we give an architecture that combines synchronous decoders with combinational decoders to carry out the decoding operations for component codes. In sequential SC decoding of polar codes, the decoder slows down every time it approaches the decision level (where decisions are made sequentially and number of parallel calculations decrease). In a hybrid-logic SC decoder, the combinational decoder is used near the decision level to speed up the SC decoder by taking advantage of the GCC structure of polar code. The GCC structure is illustrated in Fig.~\ref{fig:hybrid_enc}, which shows that a polar code $\mathcal{C}$ of length $N=8$ can be seen as the concatenation of two polar codes $\mathcal{C}_{1}$ and $\mathcal{C}_{2}$ of length $N^{\prime}=N/2=4$, each. 

The dashed boxes in Fig.~\ref{fig:hybrid_enc} represent the component codes $\mathcal{C}_{1}$ and $\mathcal{C}_{2}$. The input bits of component codes are \mbox{$\mathbf{\hat{u}}^{(1)}=(\hat{u}^{(1)}_{0}, \ldots , \hat{u}^{(1)}_{3})=(\hat{u}_{0}, \ldots , \hat{u}_{3})$} and \mbox{$\mathbf{\hat{u}}^{(2)}=(\hat{u}^{(2)}_{0}, \ldots , \hat{u}^{(2)}_{3})=(\hat{u}_{4}, \ldots , \hat{u}_{7})$}. For a polar code of block length $8$ and $R=1/2$, the frozen bits are $\hat{u}_{0}$, $\hat{u}_{1}$, $\hat{u}_{2}$, and $\hat{u}_{4}$. This makes $3$ input bits of $\mathcal{C}_{1}$ and $1$ input bit of $\mathcal{C}_{2}$ frozen bits; thus, $\mathcal{C}_{1}$ is a $R=3/4$ code with $\hat{u}^{(1)}_{0}$, $\hat{u}^{(1)}_{1}$, $\hat{u}^{(1)}_{2}$ and $\mathcal{C}_{2}$ is a $R=1/4$ code with $\hat{u}^{(2)}_{0}$ frozen. 

Encoding of $\mathcal{C}$ is done by first encoding $\mathbf{\hat{u}}^{(1)}$ and $\mathbf{\hat{u}}^{(2)}$ separately using encoders for block length $4$ and obtain coded outputs $\mathbf{\hat{x}}^{(1)}$ and $\mathbf{\hat{x}}^{(2)}$. Then, each pair of coded bits $\left(\mathbf{\hat{x}}_{i}^{(1)}, \mathbf{\hat{x}}_{i}^{(2)}\right)$, \mbox{$0 \leq i \leq 3$}, is encoded again using encoders for block length $2$ to obtain the coded bits of $\mathcal{C}$. 
\begin{figure}[hbt]
   %\centering
			\hspace{0.3cm}\includegraphics[width=3.0in,height=3.0in,keepaspectratio]{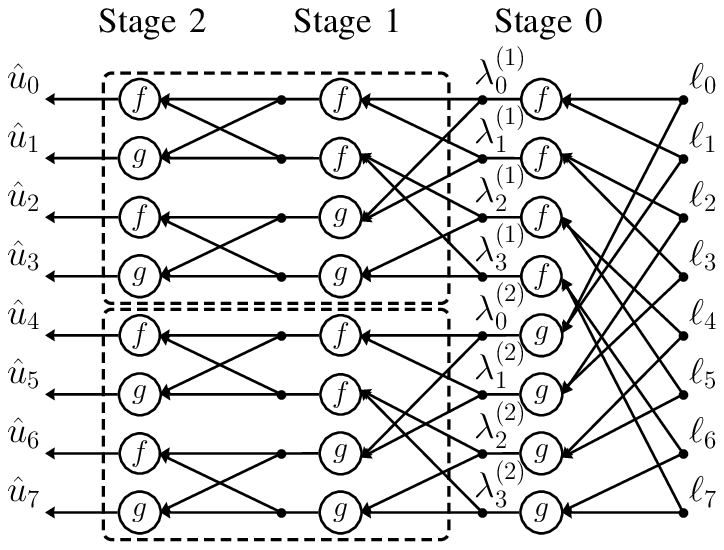}
			\vspace{0.1cm}
\caption{Decoding trellis for hybrid-logic decoder ($N=8$ and $N^{\prime}=4$)} 
\label{fig:hybrid}
\end{figure} 

Decoding of $\mathcal{C}$ is done in a reversed manner with respect to encoding explained above. Fig.~\ref{fig:hybrid} shows the decoding trellis for the given example. Two separate decoding sessions for block length $4$ are required to decode component codes $\mathcal{C}_{1}$ and $\mathcal{C}_{2}$. We denote the input LLRs for component codes as \mbox{$\boldsymbol{\lambda}^{(1)}$ and $\boldsymbol{\lambda}^{(2)}$}, as shown in Fig.~\ref{fig:hybrid}. These inputs are calculated by the operations at stage 0. The frozen bit indicator vector of $\mathcal{C}$ is \mbox{$\boldsymbol{a}=(0, 0, 0, 1, 0, 1, 1, 1)$} and the frozen bit vectors of component codes are \mbox{$\boldsymbol{a}^{(1)}=(0, 0, 0, 1)$} and \mbox{$\boldsymbol{a}^{(2)}=(0, 1, 1, 1)$}. It is seen that $\boldsymbol{\lambda}^{(2)}$ depends on the decoded outputs of $\mathcal{C}_{1}$, since $g$ functions are used to calculate $\boldsymbol{\lambda}^{(2)}$ from input LLRs. This implies that the component codes cannot be decoded in parallel.  

The dashed boxes in Fig.~\ref{fig:hybrid} show the operations performed by a combinational decoder for $N^{\prime}=4$. The operations outside the boxes are performed by a synchronous decoder. The sequence of decoding operations in this hybrid-logic decoder is as follows: a synchronous decoder takes channel observations LLRs and use them to calculate intermediate LLRs that require no partial-sums at stage $0$. When the synchronous decoder completes its calculations at stage $0$, the resulting intermediate LLRs are passed to a combinational decoder for block length $4$. The combinational decoder outputs $\hat{u}_{0}, \ldots, \hat{u}_{3}$ (uncoded bits of the first component code) while the synchronous decoder waits for a period equal to the maximum path delay of combinational decoder. The decoded bits are passed to the synchronous decoder to be used in partial-sums ($\hat{u}_{0} \oplus \hat{u}_{1} \oplus \hat{u}_{2} \oplus \hat{u}_{3}$, $\hat{u}_{1} \oplus \hat{u}_{3}$, $\hat{u}_{2} \oplus \hat{u}_{3}$, and $\hat{u}_{3}$). The synchronous decoder calculates the intermediate LLRs using these partial-sums with channel observation LLRs and passes the calculated LLRs to the combinational decoder, where they are used for decoding of $\hat{u}_{4}, \ldots, \hat{u}_{7}$ (uncoded bits of the second component code). Since the combinational decoder architecture proposed in this work can adapt to operate on any code set using the frozen bit indicator vector input, a single combinational decoder is sufficient for decoding all bits. During the decoding of a codeword, each decoder (combinational and sequential) is activated $2$ times.

Algorithm~\ref{alg:alghybrid} shows the algorithm for hybrid-logic polar decoding for general $N$ and $N^{\prime}$. For the $i^{th}$ activation of combinational and sequential decoders, \mbox{$1 \leq i \leq N/N^{\prime}$}, the LLR vector that is passed from synchronous to combinational decoder, the frozen bit indicator vector for the $i^{th}$ component code, and the output bit vector are denoted by \mbox{$\boldsymbol{\lambda}^{(i)}=(\lambda^{(i)}_{0},\ldots, \lambda^{(i)}_{N^{\prime}-1})$}, \mbox{$\boldsymbol{a}^{(i)}=(a_{(i-1)N^{\prime}}, \ldots , a_{iN^{\prime}-1})$}, and \mbox{$\mathbf{\hat{u}}^{(i)}=(\hat{u}_{(i-1)N^{\prime}}, \ldots , \hat{u}_{iN^{\prime}-1})$}, respectively. The function DECODE\_SYNCH represents the synchronous decoder that calculates the intermediate LLR values at stage $(\log _{2}(N/N^{\prime})-1)$, using the channel observations and partial-sums at each repetition. 
\begin{figure}[!t]
 \removelatexerror
  \begin{algorithm}[H]
	 \caption{$\textsc{HL\_}\textsc{Decode}(\boldsymbol{\ell}, \boldsymbol{a}, N^{\prime})$}
	 \label{alg:alghybrid}
   \For{$i = 1$ to $N/N^{\prime}$}
   {
			\eIf{$i==1$}{
				$\boldsymbol{\lambda}^{(i)} \gets \textsc{Decode\_}\textsc{Synch}(\boldsymbol{\ell}, i, N^{\prime})$ 
				}{
				$\boldsymbol{\lambda}^{(i)} \gets \textsc{Decode\_}\textsc{Synch}(\boldsymbol{\ell}, i, N^{\prime}, \mathbf{\hat{u}}^{(i-1)})$	
			}
			$\mathbf{\hat{u}}^{(i)} \gets \textsc{Decode}(\boldsymbol{\lambda}^{(i)}, \boldsymbol{a}^{(i)})$
   }
	\Return $\mathbf{\hat{u}}$
  \end{algorithm}
\end{figure}

During the time period in which combinational decoder operates, the synchronous decoder waits for $\left\lceil D_{N^{\prime}}\cdot f_{c}\right\rceil$ clock cycles, where $f_{c}$ is the operating frequency of synchronous decoder and $D_{N^{\prime}}$ is the delay of a combinational decoder for block length $N^{\prime}$. We can calculate the approximate latency gain obtained by a hybrid-logic decoder with respect to the corresponding synchronous decoder as follows: let $\mathrm{L}_{\mathrm{S}}\left(N\right)$ denote the latency of a synchronous decoder for block length $N$. The latency reduction obtained using a combinational decoder for a component code of length-$N^{\prime}$ in a single repetition is \mbox{$\mathrm{L}_{\mathrm{r}}\left(N'\right)=\mathrm{L}_{\mathrm{S}}\left(N'\right)-\left\lceil D_{N^{\prime}}\cdot f_{c}\right\rceil$}. In this formulation, it is assumed that no numerical representation conversions are needed when LLRs are passed from synchronous to combinational decoder. Furthermore, we assume that maximum path delays of combinational and synchronous decoders do not change significantly when they are implemented together. Then, the latency gain factor can be approximated as
\begin{align}
\mathrm{g}(N,N^{\prime})\approx\frac{\mathrm{L}_{\mathrm{S}}\left(N\right)}{\mathrm{L}_{\mathrm{S}}\left(N\right)-\left(N/N^{\prime}\right)\mathrm{L}_{\mathrm{r}}\left(N^{\prime}\right)}. 
\label{eq:gain}
\end{align} 
The approximation is due to the additional latency from partial-sum updates at the end of each repetition using the $N^{\prime}$ decoded bits. Efficient methods for updating partial sums can be found in \cite{anefficientpartsumnet} and \cite{ascalablesc}. This latency gain multiplies the throughput of synchronous decoder, so that:
\begin{align}
\mathrm{TP}_{\mathrm{HL}}(N,N^{\prime})=\mathrm{g}(N,N^{\prime})\: \mathrm{TP}_{\mathrm{S}}(N), \nonumber
\end{align} 
where $\mathrm{TP}_{\mathrm{S}}(N,N^{\prime})$ and $\mathrm{TP}_{\mathrm{HL}}(N)$ are the throughputs of synchronous and hybrid-logic decoders, respectively. An example of the analytical calculations for throughputs of hybrid-logic decoders is given in Section~\ref{sec:implementation}.
\subsection{Analysis}
\label{sec:analysis}
In this section, we analyze the complexity and delay of combinational architectures. We benefit from the recursive structure of polar decoders (Algorithm~\ref{alg:2NbyN_dec}) in the provided analyses.
\subsubsection{Complexity}
Combinational decoder complexity can be expressed in terms of the total number of comparators, adders, and subtractors in the design, as they are the basic building blocks of the architecture with similar complexities. 

First, we estimate the number of comparators.
Comparators are used in two different places in the combinational decoder as explained in Section~\ref{sec:combdec}: in implementing the function $f$ in \eqref{eq:f_minsum}, 
and as part of decision logic for odd-indexed bits. 
Let $c_{N}$ denote the number of comparators used for implementing the function $f$ for a decoder of block length $N$. 
From Algorithm~\ref{alg:2NbyN_dec}, we see that the initial value of $c_N$ may be taken as $c_{4}=2$.
From Fig.~\ref{fig:N4decoder}, we observe that there is the recursive relationship
$$
c_{N}=2c_{N/2}+\frac{N}{2}=2\left(2c_{N/4}+\frac{N}{4}\right)+\frac{N}{2}=\ldots.
$$
This recursion has the following (exact) solution
$$
c_N = \frac{N}{2}\log_2\frac{N}{2}
$$
as can be verified easily.

Let $s_{N}$ denote the number of comparators used for the decision logic in a combinational decoder of block length $N$. 
We observe that $s_{4}=2$ and more generally $s_N = 2s_{N/2}$;
hence,
\begin{align}
s_{N}=\frac{N}{2}. \nonumber 
\end{align}

Next, we estimate the number of adders and subtractors. The function $g$ of \eqref{eq:g_minsum} is implemented using an adder and a subtractor, as explained in Section~\ref{sec:combdec}. 
We define $r_{N}$ as the total number of adders and subtractors in a combinational decoder for block length $N$. 
Observing that $r_N = 2c_N$, we obtain
$$
r_{N}=N\log _{2}\left(N/2\right).
$$

Thus, the total number of basic logic blocks with similar complexities is given by
\begin{align}
	c_{N}+s_{N}+r_{N}=N\left(\frac{3}{2}\log _{2}\left(N\right)-1\right), 
\label{eq:complexity}
\end{align}\\
which shows that the complexity of the combinational decoder is roughly $N\log _{2}\left(N\right)$.
\subsubsection{Combinational Delay}
\label{sec:delay_analysis}
We approximately calculate the delay of combinational decoders using Fig.~\ref{fig:2NbyN_fig}. The combinational logic delays, excluding interconnect delays, of each component forming DECODE($\boldsymbol{\ell}, \boldsymbol{a}$) block is listed in Table~\ref{table:componentdelays}. 

\begin{table}[hbt!]
\caption{Combinational Delays of Components in DECODE($\boldsymbol{\ell}, \boldsymbol{a}$)} % title of Table
\centering % used for centering table
\begin{tabular}{>{\centering\arraybackslash} m{1.0in} | >{\centering\arraybackslash} m{0.5in}} % centered columns (4 columns)
\hline %inserts double horizontal lines
Block & Delay  \\  [0.5ex] % inserts table 
%heading
\hline\hline % inserts single horizontal line
$f_{N/2}(\boldsymbol{\ell})$ 	&  $\delta_{c}+\delta_{m}$	   \\  \hline
DECODE($\boldsymbol{\ell}^{\prime}, \boldsymbol{a}^{\prime}$)	&  $D^{\prime}_{N/2}$  	\\  \hline
ENCODE($\mathbf{v}$) &  $E_{N/2}$		\\   \hline
$g_{N/2}(\boldsymbol{\ell}, \mathbf{v})$ &  $\delta_{m}$		\\   \hline
DECODE($\boldsymbol{\ell}^{\prime \prime}, \boldsymbol{a}^{\prime \prime}$)	&  $D^{\prime \prime}_{N/2}$   \\   \hline
%\hline %inserts single line
\end{tabular}
\label{table:componentdelays} % is used to refer this table in the text
\end{table} 

The parallel comparator block $f_{N/2}(\boldsymbol{\ell})$ in Fig.~\ref{fig:2NbyN_fig} has a combinational delay of $\delta_{c}+\delta_{m}$, where $\delta_{c}$ is the delay of a comparator and $\delta_{m}$is the delay of a multiplexer. The delay of the parallel adder and subtractor block $g_{N/2}(\boldsymbol{\ell}, \mathbf{v})$ appears as $\delta_{m}$ due to the precomputation method, as explained in Section~\ref{sec:combdec}. The maximum path delay of the encoder can be approximated as $E_{N/2}\approx\left[\log _{2}\left(\frac{N}{2}\right)\right]\delta_{x}$, where $\delta_{x}$ denotes the propagational delay of a $2$-input XOR gate.  

We model $D^{\prime}_{N/2}\approx D^{\prime \prime}_{N/2}$, although it is seen from Fig.~\ref{fig:2NbyN_fig} that DECODE($\boldsymbol{\ell}^{\prime}, \boldsymbol{a}^{\prime}$) has a larger load capacitance than DECODE($\boldsymbol{\ell}^{\prime \prime}, \boldsymbol{a}^{\prime \prime}$) due to the ENCODE($\mathbf{v}$) block it drives. However, this assumption is reasonable since the circuits that are driving the encoder block at the output of DECODE($\boldsymbol{\ell}^{\prime}, \boldsymbol{a}^{\prime}$) are bit-decision blocks and they compose a small portion of the overall decoder block. 
Therefore, we can express $D_{N}$ as
\begin{align}
D_{N}=2D^{\prime}_{N/2}+\delta_{c}+2\delta_{m}+E_{N/2}.
\label{eq:delayeq}
\end{align}

We use the combinational decoder for $N=4$ as the base decoder to obtain combinational decoders for larger block lengths in Section~\ref{sec:combdec}. Therefore, we can write $D_{N}$ in terms of $D^{\prime}_{4}$ and substitute the expression for $D^{\prime}_{4}$ to obtain the final expression for combinational delay. Using the recursive structure of combinational decoders, we can write
\begin{align}
	D_{N}=\frac{N}{4}D^{\prime}_{4}&+\left(\frac{N}{4}-1\right)(\delta_{c}+2\delta_{m}) \nonumber \\ 
	&+\left(\frac{3N}{4}-\log _{2}\left(N\right)-1\right)\delta_{x}+\mathrm{T}_{N}.   
\label{eq:delayeq_finalpre}	
\end{align} 
Next, we obtain an expression for $D^{\prime}_{4}$ using Fig.~\ref{fig:N4decoder}. Assuming $\delta_{c} \geq 3\delta_{x}+\delta_{a}$, we can write 
\begin{align}
	D^{\prime}_{4}=3\delta_{c}+4\delta_{m}+\delta_{x}+2\delta_{a},
\label{eq:delayeq_N4}	
\end{align} 
where $\delta_{a}$ represents the delay of an AND gate. Finally, substituting \eqref{eq:delayeq_N4} in \eqref{eq:delayeq_finalpre}, we get
\begin{align}
		D_{N}=N&\left(\frac{3\delta_{m}}{2}+\delta_{c}+\delta_{x}+\frac{\delta_{a}}{2}\right) \nonumber \\ 
		&-\left\{\delta_{c}+2\delta_{m}+\left[\log _{2}\left(N\right)+1\right]\delta_{x}\right\}+\mathrm{T}_{N},
\label{eq:delayeq_final}	
\end{align} 
for $N>4$. The interconnect delay of the overall design, $\mathrm{T}_{N}$, cannot be formulated since the routing process is not deterministic. 

We had mentioned in Section~\ref{sec:combdec} that the delay reduction obtained by precomputation in adders increases linearly with $N$. This can be seen by observing the expressions \eqref{eq:delayeq_finalpre} and \eqref{eq:delayeq_N4}. Reminding that we model the delay of an adder with precomputation by $\delta_{m}$, the first and second terms of \eqref{eq:delayeq_finalpre} contain the delays of adder block stages, both of which are multiplied by a factor of roughly $N/4$. This implies that the overall delay gain obtained by precomputation is approximately equal to the difference between the delay of an adder and a multiplexer, multiplied by $N/2$. 

The expression \eqref{eq:delayeq_final} shows the relation between basic logic element delays and maximum path delay of combinational decoders. As $N$ grows, the second term in \eqref{eq:delayeq_finalpre} becomes negligible with respect to the first term, making the maximum path delay linearly proportional to $\left(\frac{3\delta_{m}}{2}+\delta_{c}+\delta_{x}+\frac{\delta_{a}}{2}\right)$ with the additive interconnect delay term $\mathrm{T}_{N}$. Combinational architecture involves heavy routing and the interconnect delay is expected to be a non-negligible component in maximum path delay. The analytical results obtained here will be compared with implementation results in the next section. 

\section{Performance Results}
\label{sec:implementation}
In this section, implementation results of combinational and pipelined combinational decoders are presented. Throughput and hardware usage are studied both in ASIC and FPGA,
and a detailed discussion of the power consumption characteristics is given form the ASIC design. 

The metrics we use to evaluate ASIC implementations are throughput, energy-per-bit, and hardware efficiency, which are defined as
\begin{align}
	\mathrm{Throughput} &\mathrm{[b/s]} = \frac{N \mathrm{[bit]}}{D_{N}\mathrm{[sec]}},	\nonumber \\
	\mathrm{Energy}\mathrm{-}\mathrm{per}\mathrm{-}\mathrm{bit}& \mathrm{[J/b]} = \frac{\mathrm{Power} \mathrm{[W]}}{\mathrm{Throughput} \mathrm{[b/s]}}, \nonumber \\
	\mathrm{Hardware}  \ \mathrm{Efficiency}& \mathrm{[b/s/m^{2}]} = \frac{\mathrm{Throughput} \mathrm{[b/s]}}{\mathrm{Area} \mathrm{[m^{2}]}},  \nonumber \\
	\label{eqn:metrics}	
\end{align} 
respectively. These metrics of combinational decoders are also compared with state-of-the-art decoders. The number of look-up tables (LUTs) and flip-flops (FFs) in the design are studied in addition to throughput in FPGA implementations. Formulas for achievable throughputs in hybrid-logic decoders are also given in this section. 
\subsection{ASIC Synthesis Results}
\label{sec:asiciplementation}
\subsubsection{Post-Synthesis Results}
\label{sec:asic_postsynth}
\begin{figure}[h!]
   \centering
\includegraphics[width=3.5in,height=3.5in,clip,keepaspectratio]{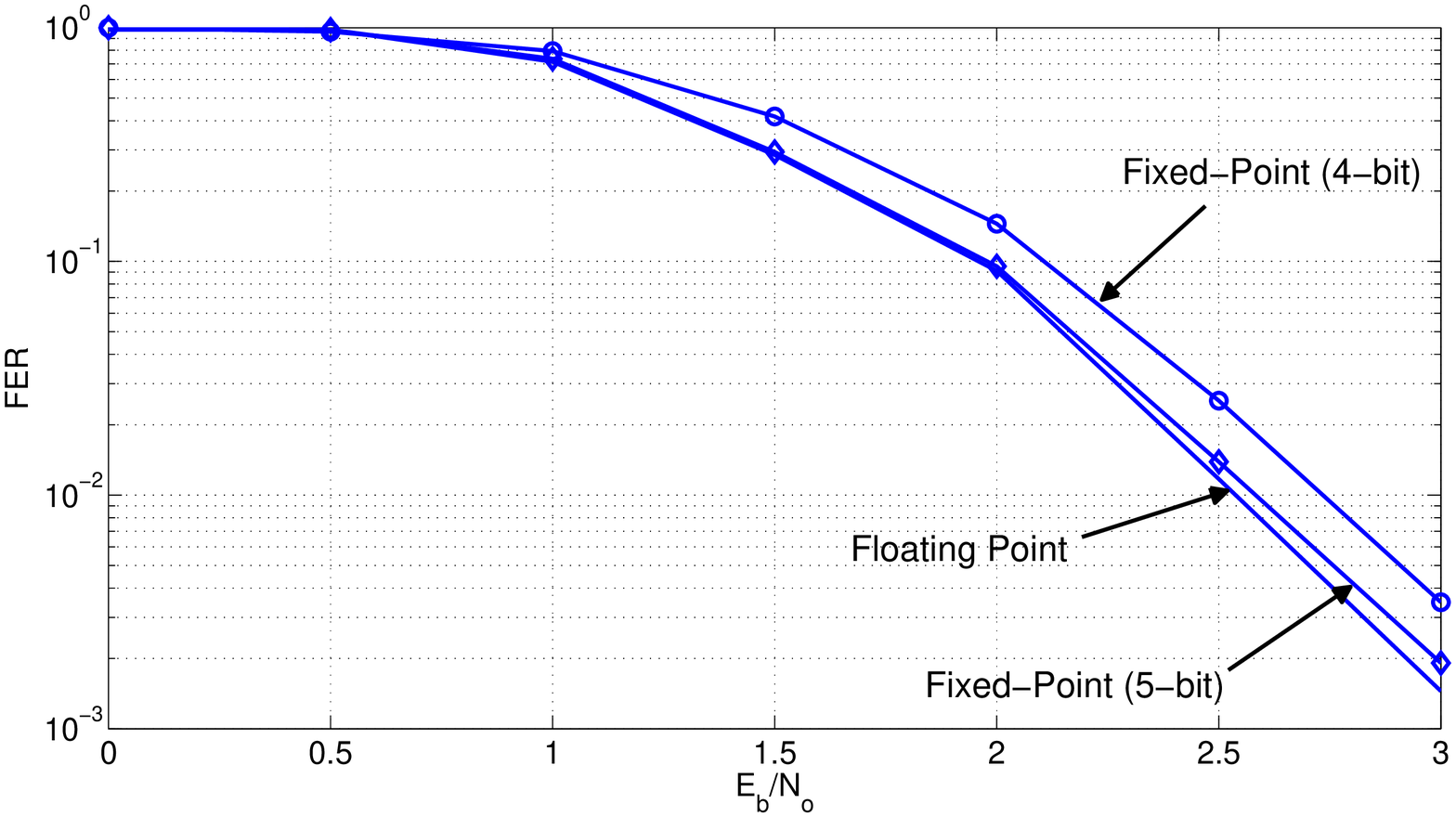}
\vspace{-0.6cm}
\caption{FER performance with different numbers of quantization bits ($N=1024$, $R=1/2$)} 
\label{fig:quant_fer}
\end{figure}

Table~\ref{table:asic} gives the post-synthesis results of combinational decoders using Cadence Encounter RTL Compiler for block lengths $2^{6}$ - $2^{10}$ with Faraday's UMC $90$~nm $1.3$~V FSD0K-A library. Combinational decoders of such sizes can be used as standalone decoders, \textit{e.g.}, wireless transmission of voice and data; or as parts of a hybrid-logic decoder of much larger size, as discussed in Section~\ref{sec:hybrid}. We use $Q=5$ bits for quantization in the implementation. As shown in Fig.~\ref{fig:quant_fer}, the performance loss with $5$-bit quantization is negligible at $N=1024$ (this is true also at lower block lengths, although not shown here).  
\begin{table}[hbt!]
\caption{ASIC Implementation Results} % title of Table
\centering % used for centering table
\begin{tabular}{>{\centering\arraybackslash} m{1.08in} | >{\centering\arraybackslash} m{0.27in} | >{\centering\arraybackslash} m{0.27in} | >{\centering\arraybackslash} m{0.27in} | >{\centering\arraybackslash} m{0.27in} | >{\centering\arraybackslash} m{0.27in}} % centered columns (4 columns)
\hline %inserts double horizontal lines
N & $2^{6}$ & $2^{7}$ & $2^{8}$ & $2^{9}$ & $2^{10}$  \\  [0.5ex] % inserts table 
%heading
\hline\hline % inserts single horizontal line
Technology & \multicolumn{5}{c}{$90$~nm, $1.3$~V} \\  \hline
Area [$\mathrm{m}\mathrm{m}^{2}$]  & 0.153 & 0.338 & 0.759 & 1.514  & 3.213    \\  \hline
Number of Cells  & 24.3K & 57.2K & 127.5K & 260.8K  & 554.3K   \\   \hline
Dec. Power [mW]  & 99.8 & 138.8 & 158.7 & 181.4  & 190.7   \\   \hline
Frequency [MHz] & 45.5 & 22.2 & 11.0 & 5.2	& 2.5															\\   \hline
Throughput [Gb/s]  & 2.92 & 2.83 & 2.81 & 2.69 & 2.56     \\   \hline
\mbox{Engy.-per-bit} [pJ/b]  & 34.1 & 49.0 & 56.4 & 67.4 & 74.5     \\  \hline % [1ex] % [1ex] adds vertical space
\mbox{Hard. Eff.} [Mb/s/$\mathrm{m}\mathrm{m}^{2}$] & 19084 & 8372 & 3700 & 1776  & 796     \\   \hline
%\hline %inserts single line
\end{tabular}
\label{table:asic} % is used to refer this table in the text
\end{table} 

The results given in Table~\ref{table:asic} verify the analytical analyses for complexity and delay. It is expected from \eqref{eq:complexity} that the ratio of decoder complexities for block lengths $N$ and $N/2$ should be approximately $2$. This can be verified by observing the number of cells and area of decoders in Table~\ref{table:asic}. As studied in Section~\ref{sec:delay_analysis}, \eqref{eq:delayeq_finalpre} implies that the maximum path delay is approximately doubled due to the basic logic elements, and there is also a non-deterministic additive delay due to the interconnects, which is also expected to at least double when block length is doubled. The maximum delay results in Table~\ref{table:asic} show that this analytical derivation also holds for the given block lengths. 

It is seen from Table~\ref{table:asic} that the removal of registers and RAM blocks from the design keeps the hardware usage at moderate levels despite the high number of basic logic blocks in the architecture. Moreover, the delays due to register read and write operations and clock setup/hold times are discarded, which accumulate to significant amounts as $N$ increases. 

\subsubsection{Power Analysis}
\label{sec:power_analysis}
Table~\ref{table:asic} shows that the power consumption of combinational decoders tends to saturate as $N$ increases. In order to fully understand this behavior, a detailed report for power characteristics of combinational decoders is given in Table~\ref{table:asic_power}. 

\begin{table}[hbt!]
\caption{Power Consumption} % title of Table
\centering % used for centering table
\begin{tabular}{ >{\centering\arraybackslash} m{0.52in} | >{\centering\arraybackslash} m{0.25in} | >{\centering\arraybackslash} m{0.25in} | >{\centering\arraybackslash} m{0.25in} | >{\centering\arraybackslash} m{0.25in} | >{\centering\arraybackslash} m{0.25in}} % centered columns (4 columns)
\hline %inserts double horizontal lines
$N$ & $2^{6}$ & $2^{7}$ & $2^{8}$ & $2^{9}$  & $2^{10}$  \\  [0.5ex] % inserts table 
%heading
\hline\hline % inserts single horizontal line
 Stat. [nW] & 701.8 & 1198.7 & 2772.8 & 6131.2 & 14846.7    \\   \hline
 Dyn. [mW] & 99.8 & 138.8 & 158.7 & 181.3  & 190.5    \\   \hline
%\hline %inserts single line
\end{tabular}
\label{table:asic_power} % is used to refer this table in the text
\end{table} 

Table~\ref{table:asic_power} shows the power consumption in combinational decoders in two parts: static and dynamic power. Static power is due to the leakage currents in transistors when there is no voltage change in the circuit. Therefore, it is proportional to the number of transistors and capacitance in the circuit (\cite{vlsibook}). By observing the number of cells given in Table~\ref{table:asic}, we can verify the static power consumption doubling in Table~\ref{table:asic_power} when $N$ is doubled. On the other hand, dynamic power consumption is related with the total charging and discharging capacitance in the circuit and defined as
\begin{align}
		P_{\mathrm{dynamic}}=\alpha C V_{DD}^{2} f_{c}, 
\label{eq:dynamic_power}
\end{align}
where $\alpha$ represents the average percentage of the circuit that switches with the switching voltage, $C$ is the total load capacitance, $V_{DD}$ is the drain voltage, and $f_{c}$ is the operating frequency of the circuit (\cite{vlsibook}). The behavior of dynamic power consumption given in Table~\ref{table:asic_power} can be explained as follows: The total load capacitance of the circuit is approximately doubled when $N$ is doubled, since load capacitance is proportional to the number of cells in the decoder. On the other hand, operating frequency of the circuit is approximately reduced to half when $N$ is doubled, as discussed above. Activity factor represents the switching percentage of load capacitance, thus, it is not affected from changes in $N$. Therefore, the multiplication of these parameters produce approximately the same result for dynamic power consumption in decoders for different block lengths. 

The decoding period of a combinational decoder is almost equally shared by the two combinational decoders for half code length. During the first half of this period, the bit estimate voltage levels at the output of the first decoder may vary until they are stabilized. These variations cause the input LLR values of the second decoder to change as they depend on the partial-sums that are calculated from the outputs of the first decoder. Therefore, the second decoder may consume undesired power during the first half of decoding period. In order to prevent this, the partial-sums are fed to the $g_{N/2}$ block through $2$-input AND gates, the second input of which is given as low during the first half of delay period and high during the second half. This method can be recursively applied inside the decoders for half code lengths in order to reduce the power consumption further. 

We have observed that small variations in timing constraints may lead to significant changes in power consumption. More precise figures about power consumption will be provided in the future when an implementation of this design becomes available.

\subsubsection{Comparison With Other Polar Decoders}
\label{sec:comp_other_polar}
In order to have a better understanding of decoder performance, we compare the combinational decoder for $N=1024$ with three state-of-the-art decoders in Table~\ref{table:asiccomparison}. We use standard conversion formulas in \cite{turbo3gpp} and \cite{690mWLDPC} to convert all designs to $65$~nm, $1.0$~V for a fair (subject to limitations in any such study) comparison.

\begin{table}[hbt!]
\caption{Comparison with State-of-the-Art Polar Decoders} % title of Table
\centering % used for centering table
\begin{threeparttable}
\begin{tabular}{>{\centering\arraybackslash} m{0.65in}|>{\centering\arraybackslash} m{0.45in} |>{\centering\arraybackslash} m{0.43in} |>{\centering\arraybackslash} m{0.43in}  |>{\centering\arraybackslash} m{0.19in} |>{\centering\arraybackslash} m{0.19in}} 
\hline %inserts double horizontal lines
 & Comb. & \cite{scasic} & \cite{anefficientpartsumnet} & \multicolumn{2}{c}{\cite{bpasicthesis}\textsuperscript{**}}  \\  
%heading
\hline\hline % inserts single horizontal line
Decoder Type & SC & SC & SC & \multicolumn{2}{c}{BP\textsuperscript{**}}       \\  \hline
Block Length  & 1024 & 1024 & 1024  & \multicolumn{2}{c}{1024}    \\  \hline 
Technology & 90~nm & $180$~nm & 65~nm  &  \multicolumn{2}{c}{65~nm}     \\   \hline
Area [mm$^{2}$] & 3.213 & 1.71 & 0.68 &  \multicolumn{2}{c}{1.476}      \\   \hline
Voltage [V] & 1.3 & 1.3 & 1.2 & 1.0 & 0.475        												\\   \hline
Freq. [MHz]  & 2.5 & 150 & 1010 & 300 & 50 														\\   \hline
Power [mW]  & 190.7 & 67 & - & 477.5 & 18.6 																\\   \hline
TP [Mb/s]  & 2560 & 49\tnote{\textdagger} & 497 & 4676 & 779.3    												\\  \hline 
\mbox{Engy.-per-bit} [pJ/b]  & 74.5 & 1370 & - & 102.1 & 23.8							\\  \hline   
\mbox{Hard. Eff.} [Mb/s/mm$^{2}$]  & 796 & 29\textsuperscript{*} & 730\textsuperscript{*} & 3168 & 528   \\   \hline  % [1ex] % [1ex] adds vertical space
\multicolumn{6}{c}{Converted to $65$~nm, $1.0$~V}       								\\  \hline 
Area [mm$^{2}$] & 1.676  & 0.223 & 0.68 &  \multicolumn{2}{c}{1.476}      \\   \hline
Power [mW]  & 81.5  & 14.3 & - & 477.5 & 82.4																\\   \hline
TP [Mb/s]  & 3544 & 136 & 497 & 4676 & 779.3    												\\  \hline 
\mbox{Engy.-per-bit} [pJ/b]  & 23.0 & 105.2 & - & 102.1 & 105.8					\\  \hline   
\mbox{Hard. Eff.} [Mb/s/mm$^{2}$]  & 2114 & 610 & 730 & 3168 & 528   \\   \hline  % [1ex] % [1ex] adds vertical space
\end{tabular}
\begin{tablenotes}
\item \textsuperscript{*} Not presented in the paper, calculated from the presented results
\item \textsuperscript{**} Results are given for $(1024,512)$ code at $4$dB SNR
\item \tnote{\textdagger} Information bit throughput for $(1024,512)$ code
\end{tablenotes}
\end{threeparttable}
\label{table:asiccomparison} % is used to refer this table in the text
\end{table} 
As seen from the technology-converted results in Table~\ref{table:asiccomparison}, combinational decoder provides the highest throughput among the state-of-the-art SC decoders. Combinational decoders are composed of simple basic logic blocks with no storage elements or control circuits. This helps to reduce the maximum path delay of the decoder by removing delays from read/write operations, setup/hold times, complex processing elements and their management. Another factor that reduces the delay is assigning a separate logic element to each decoding operation, which allows simplifications such as the use of comparators instead of adders for odd-indexes bit decisions. Furthermore, the precomputation method reduces the delays of addition/subtraction operations to that of multiplexers. These elements create an advantage to the combinational decoders in terms of throughput with respect to even fully-parallel SC decoders; and therefore, \cite{scasic} and \cite{anefficientpartsumnet}, which are semi-parallel decoders with slightly higher latencies than fully-parallel decoders. The reduced operating frequency gives the combinational decoders a low power consumption when combined with simple basic logic blocks, and the lack of read, write, and control operations. 

The use of separate logic blocks for each computation in decoding algorithm and precomputation method increase the hardware consumption of combinational decoders. This can be observed by the areas spanned by the three SC decoders. This is an expected result due to the trade-off between throughput, area, and power in digital circuits. However, the high throughput of combinational decoders make them hardware efficient architectures, as seen in Table~\ref{table:asiccomparison}. 

Implementation results for BP decoder in \cite{bpasicthesis} are given for operating characteristics at $4$~dB SNR, so that the decoder requires $6.57$ iterations per codeword for low error rates. The number of required iterations for BP decoders increase at lower SNR values Therefore, throughput of the BP decoder in \cite{bpasicthesis} is expected to decrease while its power consumption increases with respect to the results in Table~\ref{table:asiccomparison}. On the other hand, SC decoders operate with the same performance metrics at all SNR values since the total number of calculations in conventional SC decoding algorithm is constant ($N \log_{2} N$) and independent from the number of errors in the received codeword. 

The performance metrics for the decoder in \cite{bpasicthesis} are given for low-power-low-throughput and high-power-high-throughput modes. The power reduction in this decoder is obtained by reducing the operating frequency and supply voltage for the same architecture, which also leads to the reduction in throughput. Table~\ref{table:asiccomparison} shows that the throughput of the combinational decoder is only lower than the throughput of \cite{bpasicthesis} when it is operated at high-power mode. In this mode, \cite{bpasicthesis} provides a throughput which is approximately $1.3$ times larger than the throughput of combinational decoder, while consuming $5.8$ times more power. The advantage of combinational decoders in power consumption can be seen from the energy-per-bit characteristics of decoders in Table~\ref{table:asiccomparison}. The combinational decoder consumes the lowest energy per decoded bit among the decoders in comparison. 

\subsubsection{Comparison With LDPC Decoders}
\label{sec:comp_ldpc_dec}
A comparison of combinational SC polar decoders with state-of-the-art LDPC decoders is given in Table~\ref{table:comb_vs_ldpc}. The LDPC decoder presented in \cite{ldpc} is a multirate decoder capable of operating with $4$ different code rates. The LDPC decoder in \cite{parkthesis} is a high throughput LDPC decoder. It is seen from Table~\ref{table:comb_vs_ldpc} that the throughputs of LDPC decoders are higher than that of combinational decoders for $5$ and $10$ iterations without early termination. The throughput is expected to increase for higher and decrease for lower SNR values, as explained above. Power consumption and area of the LDPC decoders is seen to be higher than those of the combinational decoder. 
\begin{table}[hbt!]
\caption{Comparison with State-of-the-Art LDPC Decoders} % title of Table
\centering % used for centering table
\begin{threeparttable}
\begin{tabular}{>{\centering\arraybackslash} m{1.05in}|>{\centering\arraybackslash} m{0.55in} |>{\centering\arraybackslash} m{0.55in}  |>{\centering\arraybackslash} m{0.55in}} 
\hline %inserts double horizontal lines
 & Comb.\textsuperscript{**} & \cite{ldpc}\textsuperscript{*} & \cite{parkthesis}    \\  
%heading
\hline\hline % inserts single horizontal line
Code/Decoder Type & Polar/SC & LDPC/BP  & LDPC/BP  \\  \hline
Block Length  & 512 & 672  &  672  						\\  \hline 
Code Rate & Any & 1/2, 5/8, 3/4, 7/8 	& 1/2		\\  \hline 
Area [mm$^{2}$] & 0.79 & 1.56	&  1.60						\\   \hline
Power [mW]  & 77.5 & 361\tnote{\textdagger} 	& 782.9\tnote{\textdagger\textdagger}							\\   \hline
TP [Gb/s]  & 3.72 & 5.79\tnote{\textdagger} 	& 9.0\tnote{\textdagger\textdagger} 		\\  \hline 
\mbox{Engy.-per-bit} [pJ/b]  & 20.8 & 62.4 	& 89.5\textsuperscript{**}			\\  \hline   
\mbox{Hard. Eff.} [Gb/s/mm$^{2}$]  & 4.70 & 3.7  & 5.63\textsuperscript{**}\\   \hline  % [1ex] % [1ex] adds vertical space
\end{tabular}
\begin{tablenotes}
\item \textsuperscript{*} Technology=$65$~nm, $1.0$~V
\item \textsuperscript{**} Technology converted to $65$~nm, $1.0$~V
\item \tnote{\textdagger} Results are given for $(672,588)$ code and $5$ iterations without early termination
\item \tnote{\textdagger\textdagger} Results are given for $(672,336)$ code and $10$ iterations without early termination
\end{tablenotes}
\end{threeparttable}
\label{table:comb_vs_ldpc} % is used to refer this table in the text
\end{table} 
 
An advantage of combinational architecture is that it provides a flexible architecture in terms of throughput, power consumption, and area by its pipelined version. One can increase the throughput of a combinational decoder by adding any number of pipelining stages. This increases the operating frequency and number of registers in the circuit, both of which increase the dynamic power consumption in the decoder core and storage parts of the circuit. The changes in throughput and power consumption with the added registers can be estimated using the characteristics of the combinational decoder. Therefore, combinational architectures present an easy way to control the trade-off between throughput, area, and power. FPGA implementation results for pipelined combinational decoders are given in the next section. 
\subsection{FPGA Implementation Results}
\label{sec:fpga_implementations}
Combinational architecture involves heavy routing due to the large number of connected logic blocks. This increases hardware resource usage and maximum path delay in FPGA implementations, since routing is done through pre-fabricated routing resources as opposed to ASIC. In this section, we present FPGA implementations for the proposed decoders and study the effects of this phenomenon. 

Table~\ref{table:virtex6} shows the place-and-route results of combinational and pipelined combinational decoders on Xilinx Virtex-6-XC6VLX550T ($40$~nm) FPGA core. The implementation strategy is adjusted to increase the speed of the designs. We use RAM blocks to store the input LLRs, frozen bit indicators, and output bits in the decoders. FFs in combinational decoders are used for small logic circuits and fetching the RAM outputs, whereas in pipelined decoder they are also used to store the input LLRs and partial-sums for the second decoding function (Fig.~\ref{fig:2NbyN_fig}). It is seen that the throughputs of combinational decoders in FPGA drop significantly with respect to their ASIC implementations. This is due to the high routing delays in FPGA implementations of combinational decoders, which increase up to $90\%$ of the overall delay. 

Pipelined combinational decoders are able to obtain throughputs on the order of Gb/s with an increase in the number FFs used. Pipelining stages can be increased further to increase the throughput with a penalty of increasing FF usage. The results in Table~\ref{table:virtex6} show that we can double the throughput of combinational decoder for every $N$ by one stage of pipelining as expected. 

\begin{figure}[h!]
   \centering
	\vspace{-0.3cm}
\includegraphics[width=3.5in,height=3.5in,clip,keepaspectratio]{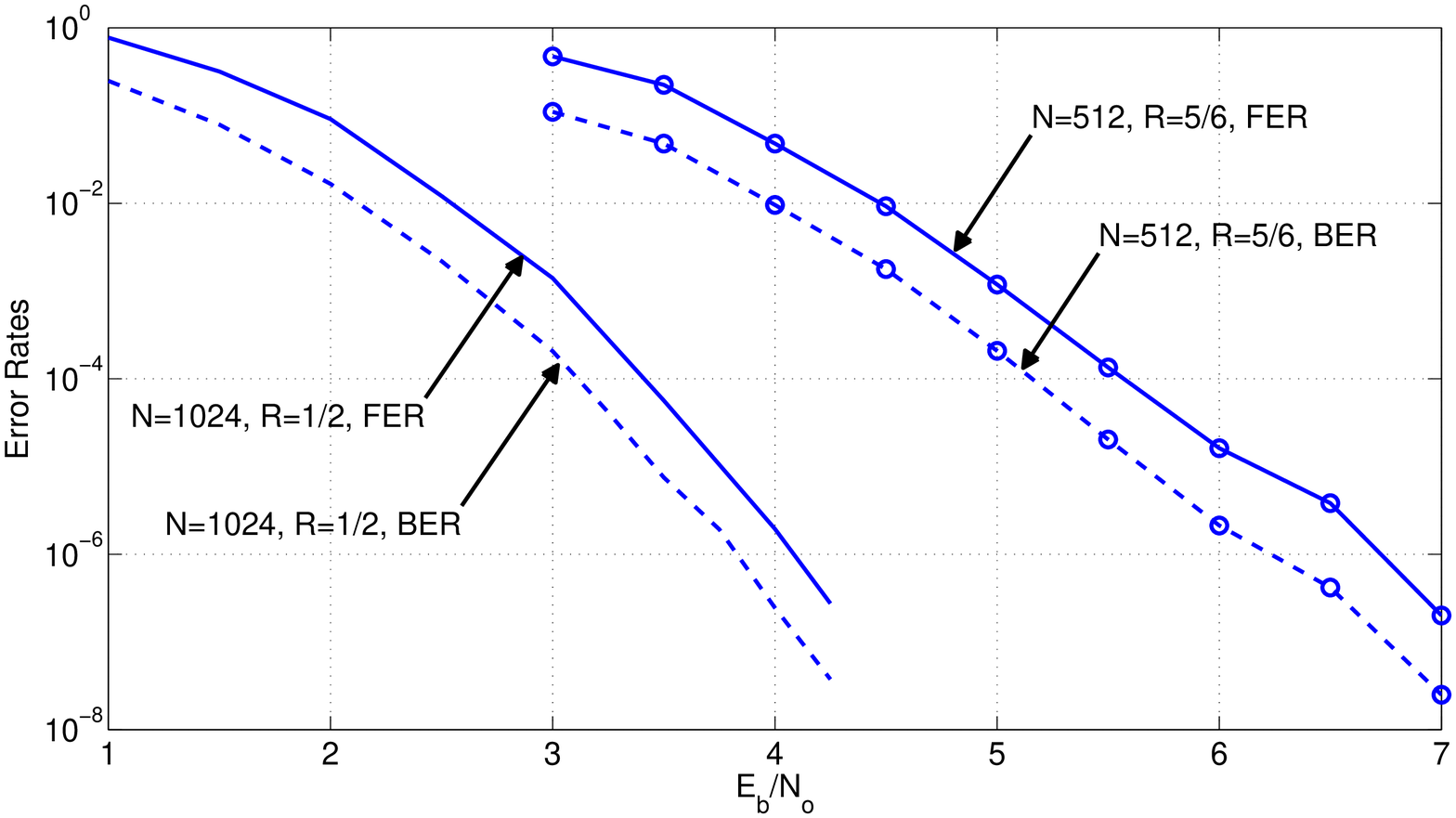}
\vspace{-0.6cm}
\caption{FER performance of combinational decoders for different block lengths and rates} 
\label{fig:floors}
\end{figure}

The error rate performance of combinational decoders is given in Fig.~\ref{fig:floors} for different block lengths and rates. The investigated code rates are commonly used in various wireless communication standards (\textit{e.g.,} WiMAX, IEEE 802.11n). It is seen from Fig.~\ref{fig:floors} that the decoders can achieve very low error rates without any error floors. 

\subsection{Throughput Analysis for Hybrid-Logic Decoders}
\label{sec:hybridresults}
As explained in Section~\ref{sec:hybrid}, a combinational decoder can be combined with a synchronous decoder to increase its throughput by a factor $g(N,N')$ as in \eqref{eq:gain}. 
In this section, we present analytical calculations for the throughput of a hybrid-logic decoder. We consider the semi-parallel architecture in \cite{asemiparallel} as the synchronous decoder part and use the implementation results given in the paper for the calculations. 

A semi-parallel SC decoder employs $P$ processing elements, each of which are capable of performing the operations \eqref{eq:f_minsum} and \eqref{eq:g_minsum} and perform one of them in one clock cycle. The architecture is called semi-parallel since $P$ can be chosen smaller than the numbers of possible parallel calculations in early stages of decoding. The latency of a semi-parallel architecture is given by
\begin{align}
	\mathrm{L}_{\mathrm{SP}}\left(N,P\right)=2N+\frac{N}{P}\log _{2}\left(\frac{N}{4P}\right). 
\label{eqn:splatency}
\end{align}
The minimum latency that can be obtained with the semi-parallel architecture by increasing hardware usage is $2N-2$, the latency of a conventional SC algorithm, when $P=N/2$. Throughput of a semi-parallel architecture is its maximum operating frequency divided by its latency. Therefore, using $N/2$ processing elements does not provide a significant multiplicative gain for the throughput of the decoder.

We can approximately calculate the approximate throughput of a hybrid-logic decoder with semi-parallel architecture using the implementation results given in \cite{asemiparallel}. Implementations in \cite{asemiparallel} are done using Stratix IV FPGA, which has a similar technology with Virtex-6 FPGA used in this work. Table~\ref{table:latencygain1} gives these calculations and comparisons with the performances of semi-parallel decoder.
\begin{table*}[hbt]
\caption{FPGA Implementation Results} % title of Table
\centering % used for centering table
\begin{tabular}{>{\centering\arraybackslash} m{0.3in} |>{\centering\arraybackslash} m{0.55in}|>{\centering\arraybackslash} m{0.55in}|>{\centering\arraybackslash} m{0.55in}|>{\centering\arraybackslash} m{0.55in} |>{\centering\arraybackslash} m{0.55in}|>{\centering\arraybackslash} m{0.55in}|>{\centering\arraybackslash} m{0.55in} |>{\centering\arraybackslash} m{0.55in} |>{\centering\arraybackslash} m{0.55in}} 
\hline 
\multirow{3}{*}{N} & \multicolumn{4}{c|}{Combinational Decoder} & \multicolumn{5}{c}{Pipelined Combinational Decoder} \\  [0.5ex]
\cline{2-10}
 & LUT & FF & RAM (bits) & TP [Gb/s] & LUT & FF & RAM (bits) & TP [Gb/s] & TP Gain \\  [0.5ex] % inserts table 
%heading
\hline\hline % inserts single horizontal line
$2^{4}$ & 1479 & 169 & 112 & 1.05 & 777 & 424 & 208 & 2.34 & 2.23   \\  \hline
$2^{5}$ & 1918 & 206 & 224 & 0.88 & 2266 & 568 & 416 & 1.92 & 2.18   \\   \hline
$2^{6}$ & 5126 & 392 & 448 & 0.85 & 5724 & 1166 & 832 & 1.80 & 2.11   \\   \hline
$2^{7}$ & 14517 & 783 & 896 & 0.82 & 13882 & 2211 & 1664 & 1.62 & 1.97   \\   \hline
$2^{8}$ & 35152 & 1561 & 1792 & 0.75 & 31678 & 5144 & 3328 & 1.58 & 2.10   \\   \hline
$2^{9}$ & 77154 & 3090 & 3584 & 0.73 & 77948 & 9367 & 6656 & 1.49 & 2.04   \\  \hline 
$2^{10}$ & 193456 & 6151 & 7168 & 0.60 & 190127 & 22928 & 13312 & 1.24 & 2.06   \\  \hline % [1ex] % [1ex] adds vertical space
%\hline %inserts single line
\end{tabular}
\label{table:virtex6} % is used to refer this table in the text
\end{table*}

Table~\ref{table:latencygain1} shows that throughput of a hybrid-logic decoder is significantly better than the throughput of a semi-parallel decoder. It is also seen that the multiplicative gain increases as the size of the combinational decoder increases. This increase is dependent on $P$, as $P$ determines the decoding stage after which the number of parallel calculations become smaller than the hardware resources and causes the throughput bottleneck. It should be noted that the gain will be smaller for decoders that spend less clock cycles in final stages of decoding trellis, such as \cite{twophase} and \cite{lowlatencysc}. The same method can be used in ASIC to obtain a high increase in throughput. 

Hybrid-logic decoders are especially useful for decoding large codewords, for which the hardware usage is high for combinational architecture and latency is high for synchronous decoders.
\begin{table}[ht]
\caption{Approximate Throughput Increase for Semi-Parallel SC Decoder} % title of Table
\centering % used for centering table
\begin{tabular}{>{\centering\arraybackslash} m{0.25in}|>{\centering\arraybackslash} m{0.3in}|>{\centering\arraybackslash} m{0.3in}|>{\centering\arraybackslash} m{0.3in}|>{\centering\arraybackslash} m{0.25in}|>{\centering\arraybackslash} m{0.3in}|>{\centering\arraybackslash} m{0.35in}} % centered columns (4 columns)
\hline %inserts double horizontal lines
\multirow{2}{*}{N} & \multirow{2}{*}{P} & f & TP$ _{\mathrm{SP}}$ & \multirow{2}{*}{$N^{\prime}$} & \multirow{2}{*}{g} & TP$ _{\mathrm{HLSP}}$ \\ 
 &  & [Mhz] & [Mb/s] &  &  & [Mb/s] \\ [0.5ex] % inserts table 
%heading
\hline\hline % inserts single horizontal line
$2^{10}$ & 64 & 173 & 85 & $2^{4}$ & 5.90 & 501   \\ \hline 
$2^{10}$ & 64 & 173 & 85 & $2^{5}$ & 6.50 & 552  \\ \hline 
$2^{10}$ & 64 & 173 & 85 & $2^{6}$ & 7.22 & 613  \\ \hline 
$2^{11}$ & 64 & 171 & 83 & $2^{4}$ & 5.70 & 473   \\ \hline 
$2^{11}$ & 64 & 171 & 83 & $2^{5}$ & 6.23 & 517 \\ \hline
$2^{11}$ & 64 & 171 & 83 & $2^{6}$ & 7.27 & 603 \\ \hline % [1ex] % [1ex] adds vertical space
%\hline %inserts single line
\end{tabular}
\label{table:latencygain1} % is used to refer this table in the text
\end{table} 
\section{Conclusion}
\label{sec:conclusion}
In this paper, we proposed a combinational architecture for SC polar decoders with high throughput and low power consumption. 
The proposed combinational SC decoder operates at much lower clock frequencies compared to typical synchronous SC decoders
and decodes a codeword in one long clock cycle. 
Due to the low operating frequency, the combinational decoder consumes less dynamic power, which reduces the overall power consumption.     

Post-synthesis results showed that the proposed combinational architectures are capable of providing 
a throughput of approximately $2.5$~Gb/s with a power consumption of $190$~mW for a $90$~nm $1.3$~V technology.
These figures are independent of the SNR level at the decoder input. 
We gave analytical formulas for the complexity and delay of the proposed combinational decoders that verify the implementation results,
and provided a detailed power analysis for the ASIC design. 
We also showed that one can add pipelining stages at any desired depth to this architecture in order to increase its throughput at the expense of increased power consumption and hardware complexity.

We also proposed a hybrid-logic SC decoder architecture that combined the combinational SC decoder with a synchronous SC decoder so as to extend the range of applicability of the purely combinational design to larger block lengths.
In the hybrid structure, the combinational part acts as an {\sl accelerator} for the synchronous decoder in improving the throughput while keeping complexity under control. 
The conclusion we draw is that the proposed combinational SC decoders offer a fast, energy-efficient, and flexible alternative for implementing polar codes.

\section*{Acknowledgment}
This work was supported by the FP7 Network of Excellence NEWCOM\# under grant agreement~318306. The authors acknowledge O. Ar\i kan, A. Z. Alkar, and A. Atalar for the useful discussions and support during the course of this work. The authors are also grateful to the reviewers for their constructive suggestions and comments.

% \bibliographystyle{IEEEtran}
% \bibliography{IEEEabrv,letter}

\begin{IEEEbiography}[{\includegraphics[width=1in,height=1.25in,clip,keepaspectratio]{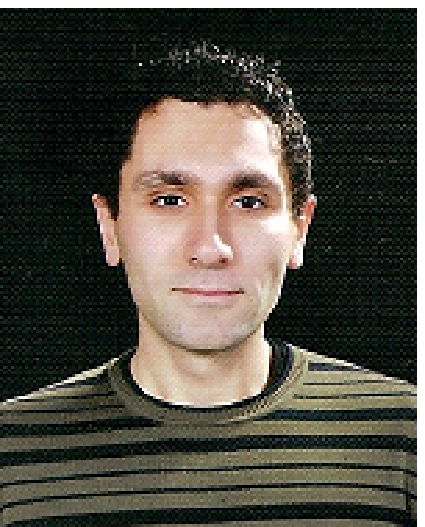}}]{Onur Dizdar}
(S'--10) was born in Ankara, Turkey, in 1986. He received the B.S. and M.S. degrees in electrical and electronics engineering from the Middle East Technical University, Ankara, Turkey
in 2008 and 2011. He is currently a Ph.D. candidate in the Department of Electrical and Electronics Engineering, Bilkent University, Ankara, Turkey. He also works as a Senior Design Engineer in ASELSAN, Turkey. 
\end{IEEEbiography} 
\begin{IEEEbiography}[{\includegraphics[width=1in,height=1.25in,clip,keepaspectratio]{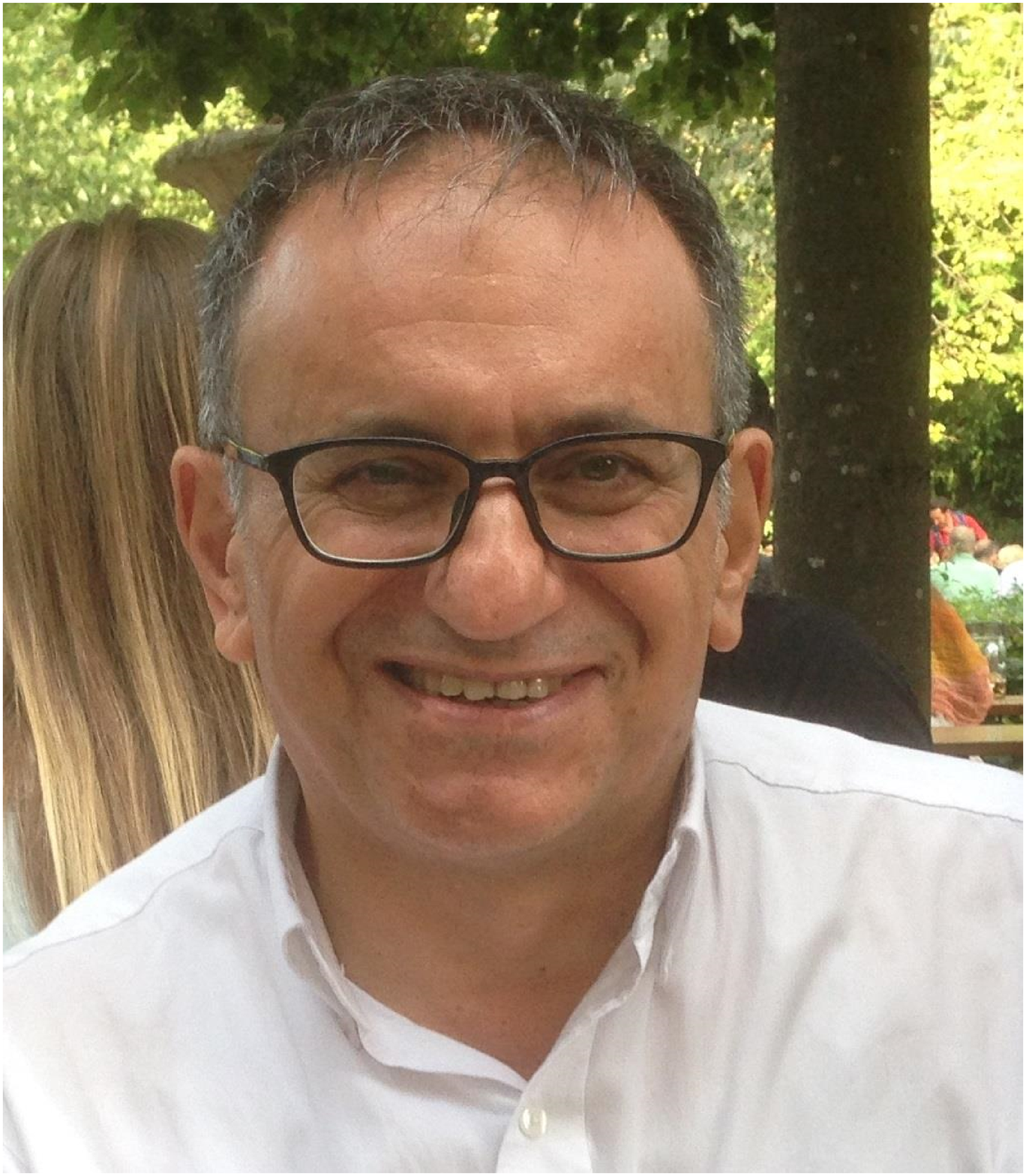}}]{Erdal Ar\i kan}
(S84), (M79), (SM94), (F11) was born in Ankara, Turkey, in 1958. He received the B.S. degree from the California Institute of Technology, Pasadena, CA, in 1981, and the S.M. and Ph.D. degrees from the Massachusetts Institute of Technology, Cambridge, MA, in 1982 and 1985, respectively, all in Electrical Engineering. Since 1987 he has been with the Electrical-Electronics Engineering Department of Bilkent University, Ankara, Turkey, where he works as a professor. He is the receipient of {\sl 2010 IEEE Information Theory Society Paper Award} and the {\sl 2013 IEEE W.R.G. Baker Award}, both for his work on polar coding.
\end{IEEEbiography}

\newpage

\end{document}